\documentclass[universe,article,publish,moreauthors]{mdpi}
\UseRawInputEncoding
\usepackage{amsmath}
\usepackage{amsfonts}
\usepackage{amssymb,bm}
\usepackage{siunitx}
\usepackage{color}
\usepackage{braket}
\usepackage{graphics}

\firstpage{1}
\pubvolume{9}
\issuenum{1}
\articlenumber{34}
\pubyear{2023}
\copyrightyear{2023}
\externaleditor{Academic Editors: Chitta Ranjan Das, Alexander S. Barabash
and Vitalii A. Okorokov}
\datereceived{22 December 2022}
\dateaccepted{29 December 2022}
\datepublished{3 January 2023}
\hreflink{https://doi.org/
{\linebreak}10.3390/universe9010034} 

\Title{How to Strengthen Constraints on Non-Newtonian Gravity from Measuring
the Lateral Casimir Force}

\TitleCitation{How to strengthen constraints on non-Newtonian gravity from measuring
lateral Casimir force}

\Author{
Galina L. Klimchitskaya ${}^{1,2}$\orcidA{},
Vladimir M. Mostepanenko ${}^{1,2,3}$\orcidB{}}

\AuthorNames{
Galina L. Klimchitskaya, Vladimir M. Mostepanenko}

\AuthorCitation{
Klimchitskaya, G.L.; Mostepanenko, V.M.}

\address{%
${}^{1}$\quad Central Astronomical Observatory at Pulkovo of the Russian Academy of Sciences, Saint Petersburg, 196140, Russia\\
${}^{2}$\quad Peter the Great Saint Petersburg
Polytechnic University, Saint Petersburg, 195251, Russia\\
${}^{3}$\quad Kazan Federal University, Kazan, 420008, Russia}

\corres{Correspondence: g.klimchitskaya@gmail.com}

\abstract{It has been known that in the nanometer interaction range the available experimental data do not exclude the Yukawa-type corrections to Newton's gravitational law which exceed the Newtonian gravitational force by many orders of magnitude. The strongest constraints on the parameters of Yukawa-type interaction in this interaction range follow from the experiments on neutron scattering and from measurements of the lateral and normal Casimir forces between corrugated surfaces. In this work, we demonstrate that by optimizing the experimental configuration at the expense of the higher corrugation amplitudes and smaller periods of corrugations it is possible to considerably strengthen the currently available constraints within the wide interaction range from 4.5 to 37~nm. We show that the maximum strengthening by more than a factor of 40 is reachable for the interaction range of 19~nm.}

\keyword{non-Newtonian gravity; lateral Casimir force; Yukawa-type interaction;
neutron scattering; Cavendish-type experiments}

\begin{document}
\section{Introduction}

Gravitation is the most universal interaction because it manifests itself for all macroscopic
bodies as well as for elementary particles and fields. Although the gravitational force is
familiar from everyday practice, the classical and quantum theories of gravitation and their
correlation with the experimental data leave much to be desired. It is common knowledge that the
construction of quantum gravity runs into insuperable difficulties. Non of the theoretical
approaches, such as string theory, or loop quantum gravity, can be considered as the
comprehensive description of gravitation. Of particular surprise is the fact that even the
classical Newton law of gravitation is tested with sufficient precision only at relatively
large separations, whereas at submicrometer distances the corrections to Newtonian gravity,
which exceed it by many orders of magnitude, are not excluded experimentally \cite{1}.
The questions arise what is the analytic form of possible corrections to Newton's law
at short separations and how they could be detected or at least constrained experimentally.

The extended Standard Model, supersymmetry and supergravity predict a lot of hypothetical
interactions  which could coexist with Newtonian gravity at submicrometer separations
(see the recent review \cite{2} for the list of these interactions, their Lagrangian
densities and forms of effective potentials). There are both spin-independent and
depending on the spin of interacting particles potentials among them. It should be
reminded that forces caused by the former act between any macroscopic bodies whereas
the latter are averaged to zero when integrated over the volumes of unpolarized bodies.

Particular attention has been given in the literature to the Yukawa-type interaction \cite{1}.
It is described by the spin-independent potential and acts between each pair of atoms belonging
to two macrobodies. After integration over their volumes with subsequent negative differentiation
with respect to the separation distance, the Newtonian gravitational force acquires some
correction.

The Yukawa-type potential is physically realized through the exchange of light scalar
particles, such as, e.g., dilaton \cite{3}, predicted in extra dimensional theories where the
additional dimensions are spontaneously compactified at the Planck energy scale of the order
of $10^{19}~$GeV. Moreover, the Newtonian gravitational force in itself gains the Yukawa-type
correction in the extra dimensional schemes where the spontaneous compactification occurs
at much lower energy scale of the order of $1~\mbox{TeV}=10^3~$GeV \cite{4,5,6,7}.
Because of this, measuring small forces between macrobodies at short separations becomes
topical as a test for predictions of fundamental physical theories.

During the last decades, different physical phenomena and associated experiments have been
used for searching the non-Newtonian gravity and constraining its parameters.
Although the corrections to Newton's law at short separations were not found, the constraints
on the parameters of corresponding potentials, including the Yukawa one, were significantly
strengthened. Thus, the stringent constraints on the Yukawa-type corrections have been
obtained from the Cavendish-type \cite{7a,8,9,10}   and E\"{o}tvos-type \cite{11,12,13}
experiments within the interaction range in excess of a few micrometers.
In the interaction range of the order of 1~nm, the most strong constraints on the parameters
of Yukawa-type interaction have been found from the experiments on neutron scattering on
solid or gaseous targets \cite{14,15,16} (see also \cite{2} for a detailed review and
\cite{17} for a recent important result).

It is notable that in the gap from nanometers to micrometers the dominant background for
searching hypothetical interactions is formed by the fluctuation-induced Casimir force
which is far in access the gravitational force within this interaction range.
First constraints on the Yukawa- and power-type corrections to Newtonian gravity were obtained
in \cite{18} and \cite{19}, respectively. In succeeding years,  a lot of precision
experiments on measuring the Casimir force and its gradient have been performed and their
results were used for obtaining stronger constraints on the Yukawa-type corrections
to Newton's gravitational law \cite{20,21,22,23,24,25,26,27,28,29,30,31} (see also
\cite{32,33} for a review).

Until recently, the strongest constraints on the Yukawa-type corrections to Newtonian
gravity in the interaction range above 1~nm were obtained from the experiments on
neutron scattering \cite{15,16}, measurements of the lateral \cite{34,35,36} and
normal \cite{37,38,39} Casimir forces between the sinusoidally corrugated surfaces,
effective Casimir pressure \cite{24,25,26,27}, and from the differential measurements
\cite{29} where the Casimir force was nullified. In the interaction range above $8~\mu$m,
the strongest constraints on the Yukawa-type interaction were found from the
gravitational experiments \cite{8,9,10,11,12,13}.

In this paper, we consider constraints on the non-Newtonian gravity of Yukawa-type
in the nanometer interaction range with account of recent strengthening achieved in the
new experiment using the neutron scattering on a silicon structure \cite{17}.
According to our results, these constraints can be further strengthened by optimizing
the experiment on measuring the lateral Casimir force between the sinusoidally corrugated
surfaces of a sphere and a plate coated with an Au layer. It is shown that by choosing
the corrugation amplitudes on a sphere and a plate equal to 33 and 90~nm, respectively,
and a corrugation period of 200~nm, it becomes possible to strengthen
the strongest current constraints by up to more than a factor of 40 over the interaction
range from 4.5 to 37~nm. Several alternative proposals on how the constraints on the
Yukawa-type interaction could be strengthened in the nanometer interaction range
are discussed.

The paper is organized as follows. In Section~2, the Yukawa-type potential is considered
and the strongest current constraints on its parameters are presented. Section~3 contains
the derivation of prospective constraints which could be obtained from the optimized
experiment on measuring the lateral Casimir force between the sinusoidally corrugated
surfaces with equal corrugation periods. The alternative proposals aimed at strengthening
the current constraints in the nanometer interaction range are analyzed in Section~4.
Our discussion is contained in Section~5, and we will finish with Section~6 devoted
to conclusions.

\section{The Yukawa-Type Potential and Current Constraints on its
Parameters in the Range from Nanometers to Micrometers}

Most commonly, the Yukawa-type interaction is considered as an addition to Newtonian
gravity. In this case the interaction energy between two point-like masses $m_1$ and
$m_2$ spaced at a distance $r$ takes the form \cite{1}
\begin{equation}
V(r)=-\frac{Gm_1m_2}{r}(1+\alpha e^{-r/\lambda}),
\label{eq1}
\end{equation}
\noindent
where $G$ is the Newtonian gravitational constant, $\alpha$ is the strength of Yukawa
interaction and $\lambda$ is its range.

If the Yukawa-type term in (\ref{eq1}) originates from the exchange of light scalar
particles of mass $m$, $\lambda$ has the meaning of the Compton wavelength of this
particle $\lambda=\hbar/(mc)$ where
$\hbar$ is the Planck constant and $c$ is the velocity of light.
In the extra dimensional theories with the low-energy compactification scale \cite{4,5,6,7},
$\lambda$ is of the order of the size of the compact manifold.

Within the interaction range up to a few nanometers, the strongest constraints on the
parameters of Yukawa-type interaction $\alpha$ and $\lambda$ follow from experiments
measuring the scattered intensity of neutrons on solid, fluid or gaseous targets \cite{39a}.
The measurement results are compared with the theoretical scattered intensity describing
interaction of neutrons with nuclei of target atoms. The measured and theoretical scattered
intensities are found in agreement in the limits of the experimental error. Then, any
contribution of the Yukawa interaction to the theoretical scattered intensity must be
restricted by the same error. This leads to the corresponding constraints  on the
Yukawa parameters $\alpha$ and $\lambda$.

\begin{figure}[!t]
\vspace*{-4.6cm}
\centerline{\hspace*{.0cm}
\includegraphics[width=6in]{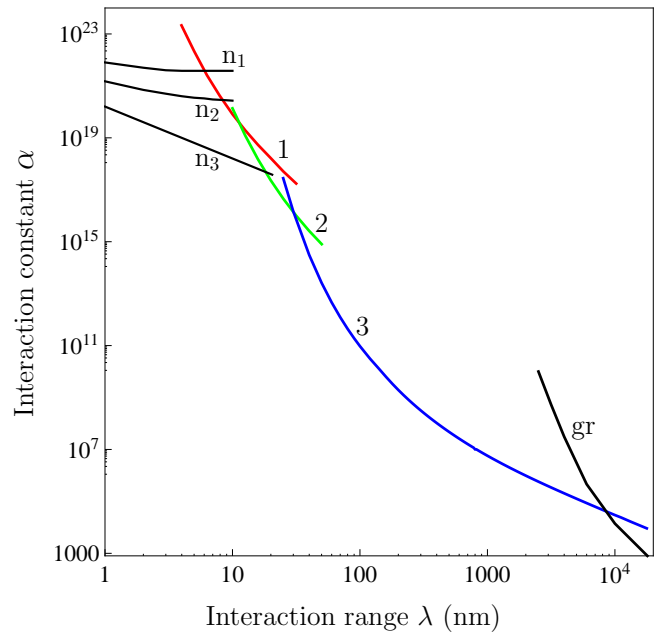}}
\vspace*{-8.5cm}
\caption{\label{fg1}
The constraints on the parameters of the Yukawa-type interaction obtained from the
experiments on neutron scattering are shown by the lines labeled n$_1$ \cite{16},
n$_2$ \cite{15}, and n$_3$ \cite{17}. The lines labeled 1, 2, and 3 show the
constraints following from measuring the lateral \cite{34,35,36} and normal
\cite{37,38,39} Casimir forces between the sinusoidally corrugated surfaces and
from the differential measurements where the Casimir force was nullified \cite{29},
respectively. The line labeled gr follows from the Cavendish-type experiments.
The regions of the ($\lambda,\alpha$)-plane above each line are excluded and below
each line are allowed.}
\end{figure}
In Figure~1, the lines labeled n$_1$, n$_2$, and n$_3$ show the constraints
obtained from measuring the scattered intensity of neutrons on xenon and helium
gases \cite{16}, atomic xenon gas \cite{15}, and on silicon \cite{17}, respectively.
For these ones and all other lines, the regions of the ($\lambda,\alpha$)-plane
above each line are excluded by the results of respective experiment whereas the
regions below each line are allowed. The constraints of the lines n$_1$, n$_2$,
and n$_3$ extend to the region $\lambda<1~$nm. In so doing at $\lambda=0.1~$nm the
lines n$_1$ and n$_2$ intersect so that for $0.03~\mbox{nm}<\lambda<0.1~$nm the
constraints of line n$_1$ \cite{16} become stronger than the constraints of
line n$_2$ \cite{15}. The strongest constraints in the region from 0.02 to
19~nm are, however, given by the line n$_3$ obtained from the recent
experiment \cite{17}.

As mentioned in Section~1, the dominant background force acting between uncharged
material bodies spaced at separation distances $a$ from a few nanometers to a few
micrometers is the Casimir force $F_C(a)$.  The Casimir force, $F_C^{\rm th}(a)$,
acting between the plane parallel structures,
can be calculated theoretically using the fundamental Lifshitz theory \cite{40,41}
(see also \cite{42,43} for a formulation in modern notations). For the bodies of
arbitrary shape $V_1$ and $V_2$, one should use the generalization of the Lifshitz
theory valid for configurations of any geometry \cite{43a,43b,44,45,45a,46,46a,46b,46c,46d}.

In precision measurements, the theoretical values of the Casimir force are usually
found to be in agreement with the mean experimental values,  $F_C^{\rm expt}(a)$,
within the limits of the total experimental error $\Delta F_C(a)$ which takes into
account both the random and systematic errors. One more force acting between the
bodies $V_1$ and $V_2$ is obtained by integration of the interaction energy (\ref{eq1})
over their volumes with subsequent negative differentiation with respect to separation
\begin{equation}
F_{\rm Yu}(a)=G\rho_1\rho_2\frac{\partial}{\partial a}\int_{V_1}\!d^3r_1
\int_{V_2}\!d^3r_2\frac{1+\alpha
e^{-|\mbox{\boldmath$r$}_1-\mbox{\boldmath$r$}_2|/\lambda}}{|\mbox{\boldmath$r$}_1
-\mbox{\boldmath$r$}_2|},
\label{eq2}
\end{equation}
\noindent
where $\rho_1$ and $\rho_2$ are the mass densities of the bodies.

Recall that in the nanometer separation range the allowable Yukawa-type corrections
exceed the Newtonian gravity by many orders of magnitudes. For this reason, one
can safely neglect by unity in the numerator of (\ref{eq2}). Thus, taking into account
that in precision experiments on measuring the Casimir force mentioned above the
Yukawa-type interaction was not observed, the constraints on its parameters can be
obtained from the inequality
\begin{equation}
|F_{\rm Yu}(a)|<\Delta F_C(a).
\label{eq3}
\end{equation}

In Figure 1, the constraints obtained following this methodology are shown by the
lines 1 and 2. The line 1 was found \cite{36} from measuring the lateral Casimir force
between the sinusoidally corrugated surfaces of a sphere and a plate coated with
Au layers \cite{34,35}. The corrugation amplitudes were equal to $A_1=85.4~$nm on
the plate and $A_2=13.7~$nm on the sphere whereas the corrugation periods were
equal to $\Lambda_1=\Lambda_2=574.7~$nm. The constraints of line 1 at different $\lambda$
were obtained \cite{36} using the measurement data at different separations:
$a_1=121.1~$nm where $\Delta F_C(a_1)=11.1~$pN,
$a_2=124.7~$nm where $\Delta F_C(a_2)=4.7~$pN, and
$a_3=137.3~$nm where $\Delta F_C(a_3)=2.5~$pN \cite{34,35}.

The constraints of line 2 in Figure 1 were found \cite{39} from measuring the normal Casimir
force between the sinusoidally corrugated surfaces of a sphere and a plate coated with
Au layers at various angles between corrugations \cite{37,38}.
In this case, the corrugation amplitudes on the plate and on the sphere were equal
to $A_1=40.2~$nm  and $A_2=14.6~$nm, respectively, and the corrugation period was
equal to $\Lambda=570.4~$nm. The angles between the axes of corrugations on both bodies
in different measurement sets were equal to $0^{\circ}$, $1.2^{\circ}$, $1.8^{\circ}$, and
$2.4^{\circ}$. The constraints of line 2 were obtained at $\theta=2.4^{\circ}$
$a=127~$nm where $\Delta F_C(a)=0.94~$pN \cite{37,38}. These constraints are the strongest
ones in the interaction range from 19 to 30~nm.

For larger $\lambda$, the strongest constraints on the Yukawa-type interaction were found
by performing the differential force measurement between an Au-coated sphere and a structured
disk consisting of Au and Si sectors coated with an Au protective layer \cite{29}.
Here, the measured quantity is the difference of forces acting on the sphere in its
positions above an Au and Si sectors. Due to the presence of an overlayer, the contributions
of the Casimir force to this difference are canceled which increases the sensitivity to the
presence of the Yukawa interaction.
In Figure~1 the obtained constraints are the strongest ones over the
interaction range from 30 to $8000~\mbox{nm}=8~\mu$m. For even larger $\lambda<4~$mm the
strongest constraints  on the parameters of Yukawa interaction follow from the Cavendish-type
experiments \cite{7a,8,9,10}. The interaction ranges of micrometers and millimeters are,
however, far beyond the subject of this paper.

\section{Prospective Constraints from Measuring  the Lateral Casimir Force
between Corrugated Surfaces}

In Section~2, we have considered the constraints on the Yukawa-type interaction obtained
from measuring the lateral Casimir force between the sinusoidally corrugated surfaces of
a sphere and a plate by means of an atomic force micriscope \cite{34,35} (see the line 1
in Figure 1). At the moment they are not the strongest ones for any $\lambda$.
In 2010, however, when these constraints have been obtained \cite{36}, they were much
more restrictive than all other constraints available at that time.

It is interesting to note in this connection that the experiment \cite{34,35} was designed
not for constraining the non-Newtonian gravity but for measuring the lateral Casimir
force in a highly nontrivial case of sufficiently small period of corrugations.
In this case, simple approximate methods for calculating the Casimir force (like the
proximity force approximation) are not applicable, and it is necessary to use the
generalization of the Lifshitz theory for the bodies of arbitrary shape based on the
scattering theory (see Section 2). With this in mind, the lateral Casimir force was
measured, compared with the exact theory, and a very good agreement was
demonstrated \cite{34,35}.

As noted in \cite{47}, the constraints on the Yukawa interaction following from
measurements of the lateral Casimir force can be strengthened by optimizing
the experimental configuration appropriately. Below we demonstrate that this
optimization in combination with an increased current precision of measurements allow
strengthening of previously found constraints \cite{36} by up to a factor of 600,
as compared to well known results \cite{15} shown by the line n$_2$ in
Figure~\ref{fg1}, and by up to more than a factor of 40 in comparison with the
most recent results shown by the line n$_3$ \cite{17}.

In our proposal of the optimized experiment on measuring the lateral Casimir force
between the sinusoidally corrugated surfaces we preserve the main features of the
already performed experiment \cite{34,35}. Specifically, we assume that the
polystyrene sphere of density $\rho_s=1.06\times 10^3~\mbox{kg/m}^3$ with the
radius $R=97.0~\mu$m is coated with a layer of chromium of thickness
$\Delta_{\rm Cr}=10~$nm and  density $\rho_{\rm Cr}=7.14\times 10^3~\mbox{kg/m}^3$
and by the outer gold layer of thickness
$\Delta_{{\rm Au},s}=50~$nm and  density $\rho_{\rm Au}=19.28\times 10^3~\mbox{kg/m}^3$.
The grating made of hard epoxy with density $\rho_g=1.08\times 10^3~\mbox{kg/m}^3$
on a Pyrex substrate of 3~mm thickness is used as a plate. The grating is coated with
an Au layer of thickness $\Delta_{{\rm Au},g}=300~$nm.

We choose only slightly larger corrugation amplitude of the grating $A_1=90~$nm,
than it was in \cite{34,35}, and by the factor of 2.4 larger corrugation amplitude on the
sphere, $A_2=33~$nm. For obtaining the equal corrugation periods, which is the necessary
condition for initiation of the lateral Casimir force, the corrugations on the sphere
should be imprinted from the grating \cite{34,35}. To produce the stronger constraints,
we choose the corrugation period of $\Lambda=200~$nm which is smaller than the one
used in the experiment by the factor of 2.9.

There is some phase shift $\varphi$ between the sinusoidal corrugations on the
sphere and on the plate. Both the lateral Casimir and lateral Yukawa-type forces
are obtained by the negative differentiation of the corresponding energy with
respect to $\varphi$
\begin{equation}
F_{C,{\rm Yu}}^{\rm lat}(a,\varphi)=-\frac{2\pi}{\Lambda}
\frac{\partial E_{C,{\rm Yu}}(a,\varphi)}{\partial\varphi}.
\label{eq4}
\end{equation}
\noindent
Note that for the corrugated surfaces the separation distance $a$ is defined between
the zeroth levels of corrugations.

The Yukawa energy $E_{\rm Yu}(a,\varphi)$ in the configuration of a corrugated sphere
and a corrugated plate coated by the chromium and gold layers as described above is
calculated by the integration of the Yukawa interaction potential over the volumes
of both bodies. The result is \cite{36}
\begin{equation}
E_{\rm Yu}(a,\varphi)=-4\pi^2G\alpha\lambda^4\Psi(\lambda)e^{-a/\lambda}
I_0\left(\frac{b(\varphi)}{\lambda}\right),
\label{eq5}
\end{equation}
\noindent
where $I_n(z)$ is the Bessel function of imaginary argument, the quantity $b(\varphi)$
is given by
\begin{equation}
b(\varphi)=\left(A_1^2+A_2^2-2A_1A_2\cos\varphi\right)^{1/2},
\label{eq6}
\end{equation}
\noindent
and the function $\Psi(\lambda)$ takes into account the layer structure of both
bodies. It is defined as
\begin{eqnarray}
&&
\Psi(\lambda)=\left[\rho_{\rm Au}-(\rho_{\rm Au}-\rho_{g})
e^{-\Delta_{{\rm Au},g}/\lambda}\right]
\label{eq7} \\
&&
\phantom{\Psi(\lambda)}\times
\left[\rho_{\rm Au}\Phi(R,\lambda)-(\rho_{\rm Au}-\rho_{\rm Cr})
\Phi(R-\Delta_{{\rm Au},s},\lambda)e^{-\Delta_{{\rm Au},s}/\lambda}\right.
\nonumber \\
&&\phantom{\Psi(\lambda)\times}
-\left.(\rho_{\rm Cr}-\rho_{s})
\Phi(R-\Delta_{{\rm Au},s}-\Delta_{\rm Cr},\lambda)
e^{-(\Delta_{{\rm Au},s}+\Delta_{\rm Cr})/\lambda}\right],
\nonumber
\end{eqnarray}
\noindent
where
\begin{equation}
\Phi(x,\lambda)\equiv x-\lambda+(x+\lambda)e^{-2x/\lambda}.
\label{eq8}
\end{equation}

Substituting (\ref{eq5}) in (\ref{eq4}), one obtains the analytic expression
for the lateral Yukawa force in the experimental configuration used to measure
the lateral Casimir force
\begin{equation}
F_{\rm Yu}^{\rm lat}(a,\varphi)=8\pi^3G\alpha\lambda^3\Psi(\lambda)e^{-a/\lambda}
\frac{A_1A_2}{b(\varphi)\Lambda}
I_1\left(\frac{b(\varphi)}{\lambda}\right)\sin\varphi.
\label{eq9}
\end{equation}

The constraints on the parameters of Yukawa interaction are obtained from the
inequality
\begin{equation}
\vphantom{\left(\frac{b(\varphi)}{\lambda}\right)}
|F_{\rm Yu}^{\rm lat}(a,\varphi)|<\Delta F_C^{\rm lat}(a),
\label{eq10}
\end{equation}
\noindent
where, in this case, $\Delta F_C^{\rm lat}(a)$ is the total experimental error in
measuring the lateral Casimir force which is the measure of agreement between the
obtained data and the exact theory.

The values of  $\Delta F_C^{\rm lat}(a)$ to be used for obtaining prospective
constraints on the Yukawa-type corrections to Newtonian gravity deserve special
attention. Taking into account larger values of the corrugation amplitudes chosen
for the prospective experiment, the minimum separation distance in this case can be
taken as $\tilde{a}_1=125~$nm (this is approximately equal to the second shortest
separation $a_2$ in the experiment \cite{34,35}). The third shortest separation
in \cite{34,35} was $a_3=137.3~$nm. This value is used as the second shortest
separation $\tilde{a}_2$ in the prospective experiment. Taking into account that
in a time passed after the  experiment \cite{34,35} was performed the measure of
agreement between measurements of the Casimir force by means of an atomic force
microscope and theory was improved by an order of magnitude \cite{48}, we put
$\Delta F_C^{\rm lat}(\tilde{a}_1)=1.11~$pN and
$\Delta F_C^{\rm lat}(\tilde{a}_2)=0.47~$pN (see Section~2). Note that an improvement
in agreement between the measurement data and theory was achieved by a significant
decrease in the residual potential difference between a sphere and a plate and use of
the cantilever of an atomic force microscope with by an order of magnitude smaller
spring constant. The latter resulted in by the factor of 10 larger calibration
constant and corresponding decrease of the systematic error in force measurements
\cite{48,48a}.

\begin{figure}[!b]
\vspace*{-7.cm}
\centerline{\hspace*{.0cm}
\includegraphics[width=6.5in]{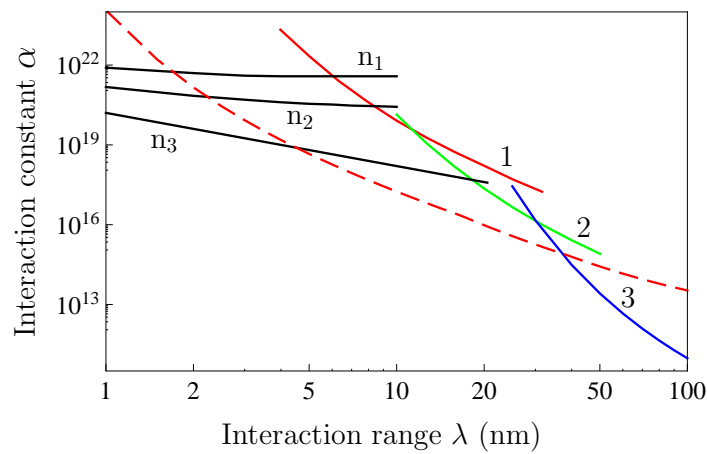}}
\vspace*{-9.1cm}
\caption{\label{fg2}
The constraints on the parameters of the Yukawa-type interaction which can be
obtained from the optimized experiment on  measuring the lateral Casimir
forces between the sinusoidally corrugated surfaces are shown by the dashed line.
The lines labeled n$_1$ \cite{16}, n$_2$ \cite{15}, and n$_3$ \cite{17}  show the
constraints following from the experiments on neutron scattering.
The lines labeled 1, 2, and 3 show the
constraints following from measurements of the lateral \cite{34,35,36} and normal
\cite{37,38,39} Casimir forces between corrugated surfaces and
from the differential measurements  \cite{29}.
The regions of the ($\lambda,\alpha$)-plane above each line are excluded and below
each line are allowed.}
\end{figure}

As a result, the prospective constraints on the Yukawa-type interaction, which can be
obtained from the optimized experiment on measuring the lateral Casimir force, are
found from (\ref{eq9}), (\ref{eq10}) with $a=\tilde{a}_1$ and $\tilde{a}_2$.
These constraints are shown by the dashed line in Figure~\ref{fg2}.
For comparison purposes, in Figure~\ref{fg2} we reproduce from  Figure~\ref{fg1}
the lines labeled n$_1$, n$_2$, and n$_3$   found
from the experiments on neutron scattering  \cite{15,16,17},
and the lines labeled 1, 2, and 3 obtained from the already performed
measurement of the lateral \cite{34,35,36} and normal
\cite{37,38,39} Casimir forces between the sinusoidally corrugated surfaces and
from the differential measurement where the Casimir force was nullified \cite{29}.

As is seen  in Figure~\ref{fg2}, the constraints shown by the dashed line are the
strongest ones in the interaction range from 4.5~nm to 37~nm. The greatest
strengthening by the factor of 41 holds at $\lambda=19~$nm. If to exclude from
comparison the recent result obtained from the neutron scattering on a silicon target
(line n$_3$)  \cite{17} and  compare the dashed line with the line  n$_2$  \cite{15},
the greatest strengthening would be by up to the factor of 600.

Thus, experiments on measuring the Casimir force have potentialities for strengthening
the constraints on the Yukawa-type corrections to Newtonian gravity and can compete on
this point with the experiments on neutron scattering. In the next section, we consider some
more suggestions recently proposed in the literature which are directed to further
strengthening of the already obtained constraints.

\section{Several Alternative Proposals}

As discussed in previous sections, the experiments on neutron scattering are actively
used for constraining the Yukawa-type corrections to Newton's gravitational law.
It was also suggested \cite{49} to constrain both the axionlike particles and the
parameters of Yukawa-type potentials by means of interferometric measurements of
neutron beams. In this case, the source beam is split in two parts which propagate
through differently oriented magnetic fields and then interfere.
Note that it was also suggested \cite{50} to use the neutron interferometry in
extra dimensional physics with the low-energy compactification scale for constraining
the Yukawa-type corrections to Newtonian gravity (see Section~1).
The compactification  radius of large extra dimensions, which is directly connected with
the parameter $\lambda$ of the Yukawa-type potential, was recently constrained
using the data sets of several neutrino experiments \cite{51}.

The spectroscopic measurements in weakly bound molecules can also be used for
constraining the Yukawa-type interaction between atoms. Thus, it was shown \cite{52}
that the measurement data of the new spectroscopic technique of photoassociation of
Yb$_2$ molecules constrain the non-Newtonian gravity in much the same manner as the
experiments on neutron scattering. It was argued \cite{52} that in combination with
the optical molecular clocks the spectroscopic measurements in weakly bound molecules
could result in up to two orders of magnitude stronger constraints on non-Newtonian
gravity than the experiments on neutron scattering. Molecular spectroscopy was
also used for constraining different kinds of potentials predicted in
various extensions of the Standard Model \cite{53}.

According to proposal of \cite{54}, the spectroscopic measurements of ionic transitions
between the Rydberg states of electronic and muonic ions lead to prospective constraints
on non-Newtonian gravity in the nanometer interaction range which are stronger than
those obtained from neutron scattering and
measuring the Casimir force. Rather wide interaction range of the
constraints which can be obtained by using the muonic ions (from $10^{-4}$ to 10~nm)
is explained in \cite{54} by the very high precision in measuring the Rydberg transitions
in the optical frequency region.

There are also several more suggestions in the literature devoted to obtaining stronger
constraints on non-Newtonian gravity at short separations. Thus, the optical setup
proposed in \cite{55} consists of a mirror and a source mass. A silica nanosphere is
trapped near the surface of a source mass. Similar to experiments on measuring the
Casimir force in the dynamic regime \cite{25,26,27}, the difference in the actual
resonance frequency of this nanosphere results from the total force gradient acting
on it. By comparing the measurement results with calculations in this configuration,
it is possible to strengthen the current constraints on the parameters of Yukawa-type
interaction. In essence, this suggestion is similar in spirit to the experiment on
measuring the thermal Casimir-Polder force between a cloud of $^{87}$Rb atoms
belonging to the Bose-Einstein condensate and a SiO$_2$ plate \cite{55a}.
In this experiment, the atomic cloud was resonantly driven into a dipole oscillation
by a magnetic field, and the measured quantity was the frequency shift caused by the
Casimir-Polder force. The constraints on the Yukawa-type interaction following from
the measurement data of \cite{55a} were obtained in \cite{36}.
According to the results of \cite{55}, the use of a nanosphere allows obtaining
 up to two orders of magnitude stronger constraints
 over the interaction range from 100~nm
to $10~\mu$m. This is, however, beyond the nanometer interaction range considered
in Section~3.

Recently the new Cavendish-type experiment was performed \cite{10,56} which is especially
optimized in order to increase possible impact of the Yukawa-type interaction.
This was achieved by substantial increasing the areas of both the source and test masses.
As  a result, up to an order of magnitude strengthening of the constraints on the
constant $\alpha$ of Yukawa-type interaction   was reached in the range from 40 to
$800~\mu$m which is not shown in Figure~\ref{fg1}.

One more suggestion for searching the non-Newtonian gravity is to perform  the
microscope E\"{o}tvos-type experiment which should test the weak equivalence principle
\cite{57}. For this purpose, it is planned to compare the low-frequency dynamics of
the free-falling test masses of cylindrical form controlled by electrostatic forces.
The expected constraints should span the interaction range from $100~\mu$m to 1~m
far away from the subject of our concern.

There are also proposals of new experiments on measuring the Casimir-Polder and
Casimir forces. Thus, it was shown \cite{58} that measurements of the Casimir-Polder
force between a Rb atom and a movable Si plate shielded by the Au film can be used
for obtaining up to two orders of magnitude stronger constraints on the Yukawa-type
corrections to Newtonian  gravity in the interaction range from a fraction of
micrometer to $10~\mu$m. Within the same interaction range, the current constraints
can be strengthened by an order of magnitude from the experiment measuring the
Casimir pressure and its gradient between two parallel plates at micrometer
separations \cite{59,60}.

In the end of this section, we mention the proposal for testing the corrections
to Newtonian gravity in the submillimeter and submicrometer ranges by using the
optical system with extremely high temporal precision \cite{61}. It remains
unknown, however, how strong constraints could be obtained by using this
experimental approach.

\section{Discussion}

In the foregoing, we have considered the corrections to Newtonian gravity
in the nanometer interaction range and different possibilities for their
strengthening. During the last few years, the strongest constraints on
non-Newtonian gravity in this interaction range were obtained from the
experiments on neutron scattering and from measurements of the lateral and
normal Casimir forces between the sinusoidally corrugated surfaces of
a sphere and a plate. Somewhat weaker constraints were found
from the experiments on measuring the Casimir force between a microsphere
coated with Au layer and a silicon carbide plate spaced at the minimum
separation of 10~nm \cite{62,63,64} and the phase shifts caused by the
Casimir-Polder interaction between Li atoms and a silicon nitride
grating \cite{64a}.

It should be noted that the recent experiment on neutron scattering
\cite{17} (see the line n$_3$ in Figures 1 and 2) has extended the
interaction range, where the neutron experiments lead to the strongest
constraints on the Yukawa-type corrections to Newtonian gravity, up to
19~nm. However, as shown in this paper, measurements of the lateral
Casimir force have a good chance to become the strongest ones in the
interaction range from 4.5 to 37~nm. This means that some kind of
competition between the experiments on neutron scattering and
measurements of the Casimir force in obtaining constraints at
nanometers is being continued. One can guess that the spectroscopic
measurements \cite{64b} will also take part in this competition.

The derivation of constraints on non-Newtonian gravity from
measurements of the Casimir force is essentially based on the
comparison between experiment and theory. It has been known, however,
that this comparison is complicated by the problem of relaxation
properties of conduction electrons. This problem lies in the fact
that theoretical predictions of the Lifshitz theory are in good
agreement with the measurement data only if the relaxation properties
of conduction electrons are omitted in computations. If, however, the
relaxation properties are taken into account, the theoretical
predictions are excluded by the data
\cite{24,25,26,27,28,29,30,42,43,48,65}. The necessity of considering
the real material properties, specific geometry of the interacting
bodies and the effect of thermal fluctuations in calculations of the
Casimir force was underlined in \cite{65a,65b,65c}. According to
\cite{65b,65c}, if the objective of some particular experiment is
to test the model of material properties, that experiment should not
be used for constraining the hypothetical interactions. During the
past ten years, however, many problems in computation of the Casimir
force mentioned in \cite{65a,65b,65c} found a solution. Thus, the
decisive differential measurement \cite{66}, where the predictions
of the Lifshitz theory with omitted and included relaxation properties
of conduction electrons differ by up to the factor of 1000, conclusively
established that for reaching an agreement with the experimental data
the computational approach should omit the relaxation properties of
conduction electrons. An active search for the theoretical
justification of this approach is in progress \cite{67,68}.
It is pertinent to note, however, that both theoretical
approaches lead to almost equal results for the lateral and normal
Casimir forces between the sinusoidally corrugated surfaces at short
separations used for obtaining the constraints on non-Newtonian
gravity in the nanometer interaction range (see the lines 1 and 2
in Figures 1 and 2). In a similar way, the mentioned distinction
between the theoretical approaches makes no impact on the results
of the Casimir-less experiment \cite{29} where the contribution of
the Casimir force is canceled (see the line 3 in Figures 1 and 2).

In spite of a notable advance which has been made in obtaining
stronger constraints on the Yukawa-type corrections to Newtonian
gravity in the nanometer interaction range, much remains to be done.
The point is that the corrections exceeding the Newton force by
up to 19 orders of magnitude are still not ruled out by the
measurement data in the range of a few nanometers. This is not an
entirely academic problem related to such fundamental unresolved
issues as, e.g., the nature of dark matter \cite{2,69}. As an
example, the Yukawa-type corrections to Newton's law make an
impact on the properties of strange quark stars. Specifically,
it is shown that observations of some astrophysical events using
the MIT bag model cannot be explained if the effects of
non-Newtonian gravity are disregarded \cite{70}. In this connection,
an investigation of different kinds of additional forces predicted
by the fundamental physics (they are often referred to by the
generic name {\it the fifth force} and the experimental limits
imposed on their parameters deserve further attention \cite{71}.

\section{Conclusions}

To conclude, the strongest current constraints on the Yukawa-type
corrections to Newton's gravitational law in the nanometer
interaction range follow from the recent experiment on neutron
scattering (see the line n$_3$ in Figure 2), from measuring the
normal Casimir force between the sinusoidally corrugated surfaces
(see the line 2 in Figure 2), and from the differential measurement
where the Casimir force is nullified (see the line 3 in the
same figure).

According to our results, these constraints can be strengthened
by up to a factor of 41 basing on the experiment measuring the
lateral Casimir force between the sinusoidally corrugated surfaces
of a sphere and a plate. We propose this experiment as a modification
of the already performed measurement of the lateral Casimir force
optimized in order to obtain the strongest constraints achievable
in this configuration. Our calculations show that the strengthening
can be obtained in the interaction range from 4.5 to 37~nm. This
would narrow the interaction region, where the strongest constraints
follow from the experiments on neutron scattering, and exclude
measurements of the normal Casimir force between the sinusoidally
corrugated surfaces from the list of experiments leading to the
strongest constraints on non-Newtonian gravity of Yukawa-type in
the nanometer interaction range.

In near future one could expect new results in this rapidly
progressing field obtained from the repetitions and modifications
of the experiments mentioned above, as well as from the alternative
experimental tests based on some other physical phenomena.

\authorcontributions{
Conceptualization, Galina L. Klimchitskaya and Vladimir M. Mostepanenko;
Investigation, Galina L. Klimchitskaya  and Vladimir M. Mostepanenko;
Writing: original draft,
Vladimir M. Mostepanenko; Writing: review \& editing, Galina L. Klimchitskaya
and Vladimir M. Mostepanenko}

\institutionalreview{Not applicable. }
\informedconsent{Not applicable. }
\dataavailability{Not applicable. }

\acknowledgments{
The work of V.M.M.~was supported by the Kazan Federal University
Strategic Academic Leadership Program.
}

\conflictsofinterest{The authors declare no conflict of interest.}

\end{paracol}
\reftitle{References}

\end{document}